**Stretch-free, shape-induced 3D Island–Bridge Networks for flexible TFTs on Silicon Planar Technology verified through Bending and Scalability to 9×9 Matrix**

*Daniel Joch*, Robert Kammel, Fabian Magerl, Mathias Rommel and Michael P.M. Jank*

Daniel Joch, Mathias Rommel, Michael P.M. Jank
Research and Development Semiconductor Devices
Fraunhofer Institute for integrated Systems and Device Technology IISB
Schottkystrasse 10, 91058 Erlangen, Germany
E-mail: daniel.joch@iisb.fraunhofer.de

Daniel Joch, Robert Kammel, Fabian Magerl, Michael P.M. Jank
Department of Electrical, Electronic and Communication Engineering, Electron Devices (LEB)
Friedrich-Alexander-Universität Erlangen-Nuremberg
Cauerstr. 6, 91058 Erlangen, Germany

ORCiD: 0000-0001-7304-0503, 0009-0001-7314-1694, 0000-0002-1141-3228, 0000-0002-6523-268

Funding: This research received no external funding.

((**Abstract text.** Maximum length 200 words. Written in the present tense.))

This study presents a CMOS-compatible, fully integrated three-dimensional island-bridge concept for flexible electronics on silicon planar technology. By embedding metal bridges within trenches in a polyimide-passivated island matrix, the approach localizes mechanical stress to the bridges whereby active components on the islands are protected from mechanical stress, enabling high-performance thin-film transistors (TFTs) on flexible substrates. A concave, arc-shape forming fill in trenches between the islands and backside etching yield freestanding 3D bridges.

Numerical simulations to determine the minimum bending radius reveal a characteristic stress distribution in the bridges during bending, with peak stresses at the bridge–island transitions. Variation of trench depth modulates von Mises stress, identifying design parameters for reliability. Experimental validation demonstrates TFT operation under bending, with stable threshold voltage, subthreshold swing, and saturation mobility across a range of bending radii; broader bridges exhibit enhanced mechanical robustness.
A 9×9 island-bridge matrix with addressable integration of TFTs across islands demonstrates the scalability of the concept. Overall, the results verify the manufacturability of stretch-free 3D metal bridges where the three-dimensional shape is defined by the topography of the concave trench filling, with integrated active devices, and confirm the mechanical and electrical functionality of the produced flexible substrates

## 1. Introduction

The integration of active circuit elements on flexible substrates is a central goal in flexible electronics. Such approaches enable installation on complex physical structures and on dynamic surfaces typical of wearable systems. Mechanical strains from bending, folding, and stretching — or combinations thereof — must be accommodated with minimal change of electrical performance.[1–3] To preserve electrical properties, hybrid electronic concepts that combine conventional inorganic materials with structural design play a key role. These concepts rely on mechanical principles, meaning that metal interconnects are configured for mechanical flexibility while active components are simultaneously protected from mechanical stress.[3–5] The structural design of hybrid electronics varies widely, and a distinction can be made between substrate engineering and the structuring of the interconnects. The ability to fabricate three-dimensional structures broadens the design space.[6,7]
Treating the substrate surface can yield wavy topographies that serve as templates for the metals deposited on top. The prepared surface acts as a shaping structure and represents a simple manufacturing route.[7,8] Common metal trace patterns include serpentine and horseshoe geometries.[4,5,9,10] Decomposing these patterns into small unit cells and recombining them enables fractal architectures, which can be extended to kirigami or origami configurations.[5,11–13] Kirigami structures, in particular, can move out of plane under mechanical stress, thus representing a transition to three-dimensional structures.[4,12]
Kim et al. were among the first to use such designs to interconnect small rigid components with metal traces, giving rise to the island-bridge concept as a matrix-like, mechanically flexible arrangement.[14,15] The concept concentrates mechanical stress in the bridge (interconnect) regions designed to deform, while islands remain as stress-free as possible to host active structural elements. This approach offers clear advantages for deploying high-performance components in flexible electronics, since established inorganic materials can serve as both conductive traces and active components.[14–16] The island-bridge concept is frequently used with three-dimensional bridge structures to extend two-dimensional patterns into an additional design dimension. The fabrication route involves transfer printing of the complete planar island-bridge assembly onto a pre-stretched, elastomeric substrate. Releasing the pre-stretch introduces compressive strain in the flexible bridges, which buckle from the surface to form a non-coplanar mesh.[17–19]
Although the island-bridge concept is promising for high stretchability with high-performance materials, the described transfer route has limitations due to its planar fabrication nature. Precise placement and alignment on the target substrate become increasingly challenging as structures scale up. Achieving a homogeneous pre-stretch distribution over the area is difficult, thereby causing failures within the matrix and reducing yield. As an alternative, direct processing on pre-stretched polymer substrates could be pursued, though this is not feasible with conventional planar techniques.[8,20]
Thus, a manufacturing route is desirable that fully integrates three-dimensional metal interconnects into silicon-based planar technology. CMOS-compatible two-dimensional island-bridge structures already demonstrated baseline fabrication capabilities, combining front- and back-end processes to produce flexible substrates.[21,22] Going beyond this, we developed in our previous work a distinct technology that, in contrast to other approaches, requires no pre-stretching of the substrate. We introduced a process flow in which three-dimensional bridges are formed into trenches using a shape-forming step and conventional etching. Based on an island-structured silicon wafer, we demonstrated a stretch-free manufacturing route for a three-dimensional island-bridge concept fully integrated into standard planar technology.[23,24] **Figure 1** provides an overview of our previous development.

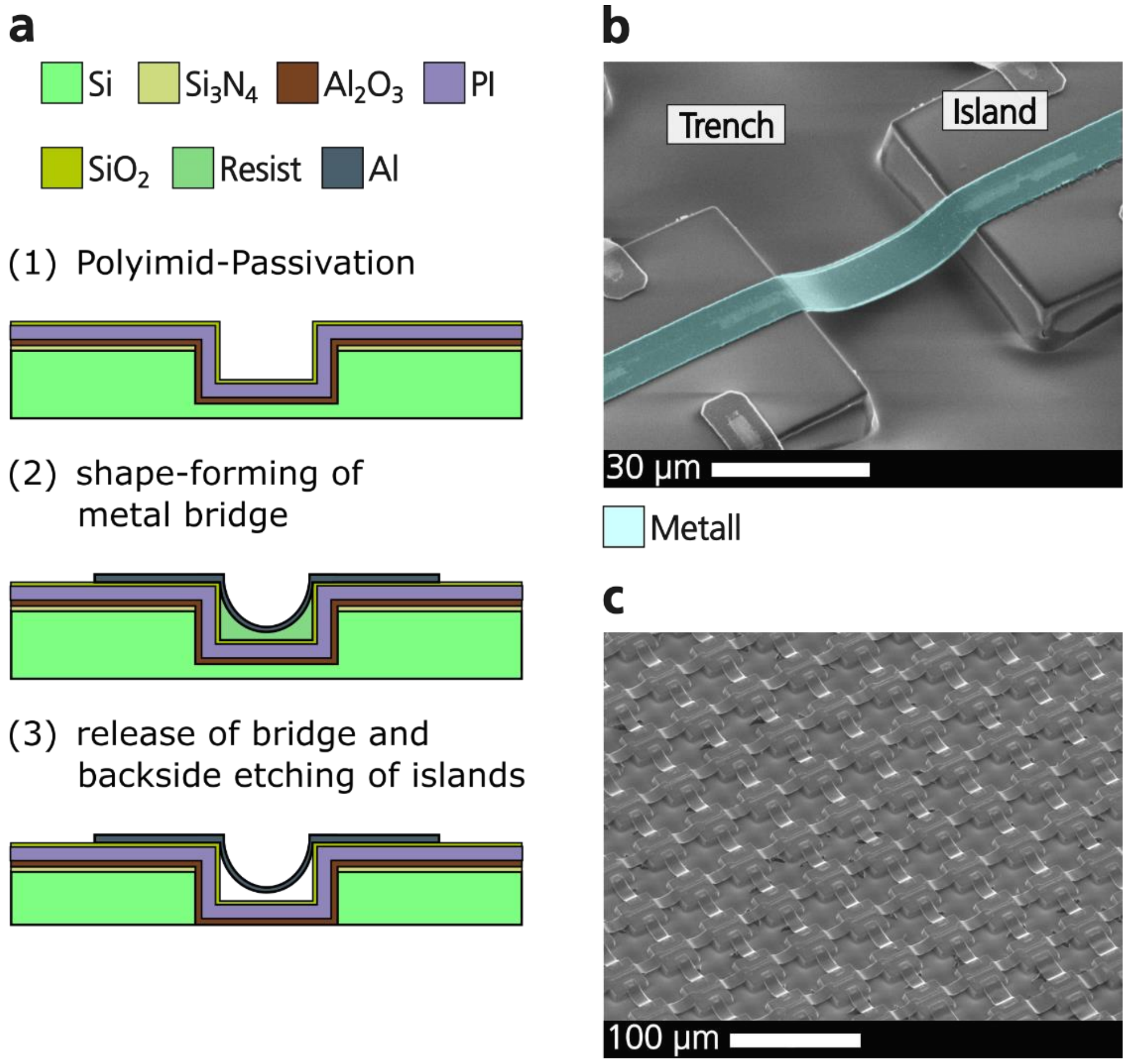

**Figure 1.** Overview of stretch-free manufacturing route of three-dimensional island-bridge concept within the silicon planar technology: a) Illustration of the three core modules. b) SEM image of a metal bridge (width 10 µm) crossing a polyimide-coated trench (depth 20 µm, width 40 µm). c) SEM image showing the island-matrix connected vertically and horizontally.[24]

The core of the process is shown in **Figure 1a**. After patterning the three-dimensional islands on the surface, polyimide serves both as passivation for potential device structures and as the flexible carrier substrate. In the trenches, a temporary polymer fill creates a shaping surface on which metallization is deposited, followed by metal patterning with a resist mask and etching. Finally, the fill is removed by cleaning in a bath of a specific solvent, and the islands are released by backside etching. The polyimide's multifunctionality is evident as the islands are embedded and render the entire substrate freestanding.[24] **Figure 1b** shows a freestanding metal bridge between two islands after processing, and **Figure 1c** illustrates the matrix of islands interconnected with their neighbors. Our previous work established this technology platform and proved the fundamental feasibility of the concept — but active devices had not yet been integrated.[23,24]

The aim of the work presented here is now to address exactly this challenge: extending the island–bridge process towards advanced, flexible thin-film electronics by integrating active devices on the islands and assessing their performance under mechanical load. Thin-film transistors (TFTs) were fabricated on each island prior to executing the core modules in **Figure 1a**. The flexible structures were subjected to bending tests across different radii, with support from numerical simulations. We also illustrate the scalability of the concept in a matrix arrangement like **Figure 1c**, where each TFT is addressable.

## 2. Results and Discussion

### 2.1. Numerical Simulation for Evaluation of the bending radius in dependence of stress distribution

To provide insight into the mechanical response of the structure under bending, numerical simulations were conducted to examine the stress distribution as a function of bending load. In this way, it is possible to determine the maximum stress associated with each bending radius and, by varying the trench depth $D$, the influence of this parameter on stress optimization can be assessed. The distance between the island was always held constant with 40 µm.
**Figure 2** shows a typical result of the numerical modeling using a defined bending radius $R_{bending}$. The geometry variation has a trench depth $D$ of 10 µm. As shown in **Figure 2a**, the bending moment was adjusted to fit the geometry to the supporting surface. The results in **Figure 2b** and **2d** show the von Mises stress $\sigma_{VM}$ in the middle of the bridge and at the transition to the island, respectively. The color gradient clearly indicates a neutral plane at mid-thickness and a maximum stress at the top and bottom edges. This is the typical stress distribution during bending of a structure, which induces compressive and tensile stresses on opposite sides. **Figure 2c** shows the normal stress $\sigma_N$ in the middle of the bridge to illustrate the typical positive and negative values arising from bending.

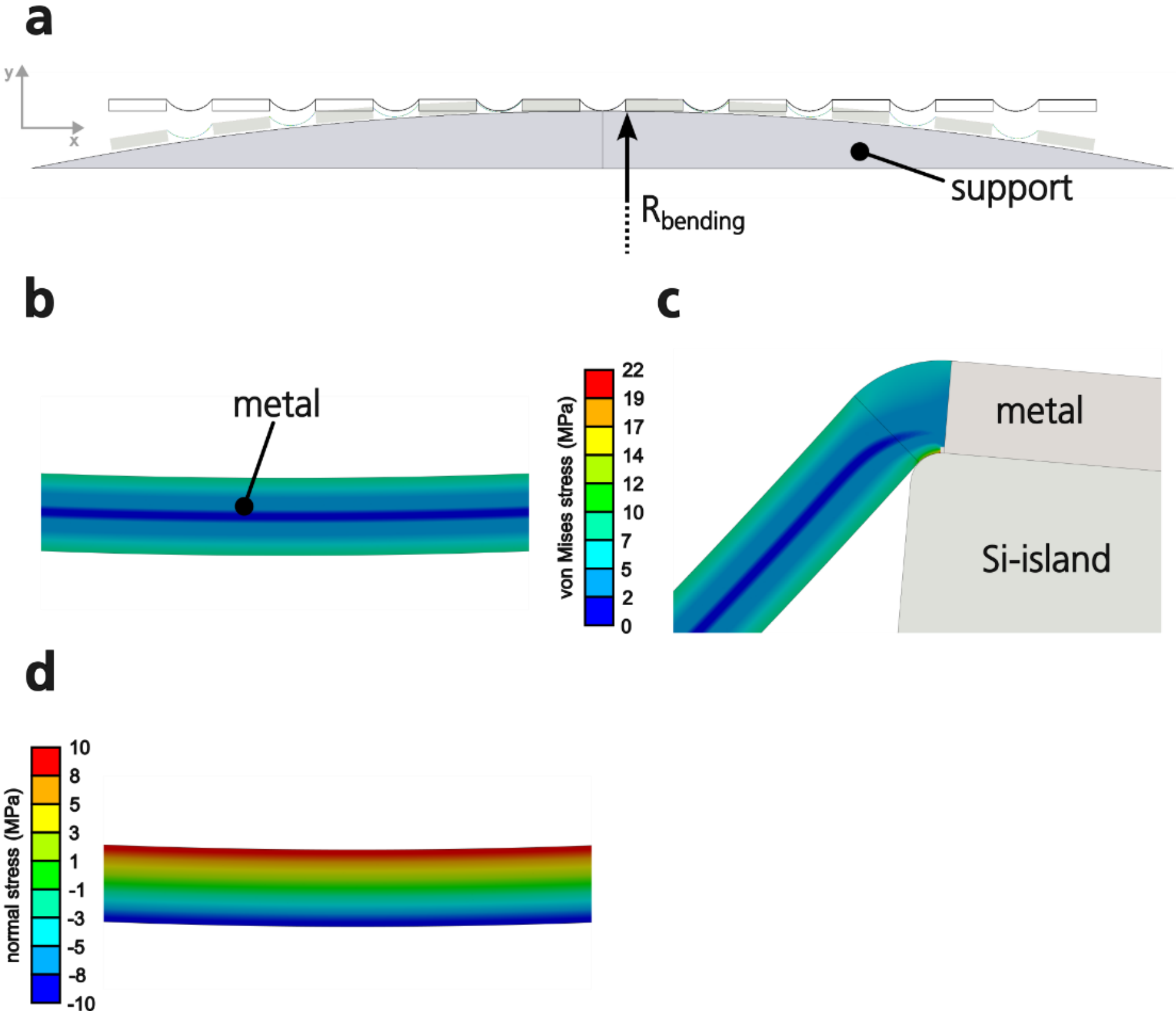

**Figure 2.** a) Schematic of the bending simulation with the start and end positions of the geometry (trench depth D = 10 µm). b) Distribution of $\sigma_{VM}$ in the middle of the metal bridge. c) $\sigma_{VM}$ at the transition to the island. d) Distribution of $\sigma_N$ illustrating the characteristic positive and negative values due to bending.

**Table 1** summarizes the extracted maximum values for each modeled bending radius. This allows for a clear comparison of the stress values with the yield strength $\gamma_{ys}$, which is introduced together with the material model in the Supporting Information. This characteristic material parameter defines the boundary between elastic and plastic deformation.[25] The latter can cause irreversible damage and must therefore be avoided. Within the material model $\gamma_{ys}$ was determined to be 545 MPa. In **Table 1** higher values for $\sigma_{VM}$ at the transition to the island can be seen, indicating that this region should be regarded as more critical with respect to $\gamma_{ys}$. This is evident from the maximum stress evaluated at the transition relative to the rest of the bridge.

**Table 1.** Evaluation of the maximum stresses in the bridge and at the transition to the island for different bending radii and trench depth D.

| | *D* = 10 µm | | *D* = 20 µm | |
|---|---|---|---|---|
| Radius $R_{bending}$ (mm) | $\sigma_{VM}$ bridge (MPa) | $\sigma_{VM}$ transition (MPa) | $\sigma_{VM}$ bridge (MPa) | $\sigma_{VM}$ transition (MPa) |
| 15 | 4.3 | 7.4 | 3.9 | 6.1 |
| 5 | 12.3 | 21.6 | 11.4 | 17.8 |
| 2 | 32.8 | 57.5 | 30.5 | 47.4 |
| 1 | 66.3 | 115.8 | 58.8 | 90.5 |

Regarding the bending radius, the values for $\sigma_{VM}$ increase with decreasing $R_{bending}$, as expected. The differences between the geometries highlight the influence of the trench depth. In our previous work on tensile loading of the structure, we showed that *D* also influences the maximum stresses in the bridge.[24] Here we can now see a similar trend, so the results confirm the statement that *D* can be regarded as a key parameter in the design of these bridges. A comparison of the two geometries is shown in **Figure 3**, illustrating the critical region during bending with a radius of 1 mm. Keeping the same range for the stress in the legend, the color gradient indicates lower values for *D* = 20 µm, particularly evident at the structure edges.
Note that even at a radius of 1 mm, the achieved stress remains well below the $\gamma_{ys}$ of 545 MPa, which was determined by literature, indicating that plastic deformation is unlikely. Beyond a certain point, plastic deformation would become irreversible and could cause permanent damage. To verify this, the plastic strain $\varepsilon_{pl}$ was extracted from the simulation for each geometry and radius. For the radius of 1 mm, $\varepsilon_{pl}$ = 0.16% for *D* = 10 µm and $\varepsilon_{pl}$ = 0.12% for *D* = 20 µm were observed. In the supporting information, the material model used for the simulation is described in terms of yield strength and critical plastic strain. There, also based on literature research, damage to the structure is assumed when $\varepsilon_{pl}$ exceeds a value of 1%. The strains occurring here are therefore well below this limit and were observed in only a single element at the corner of the transition, where a singular elevation cannot be ruled out.

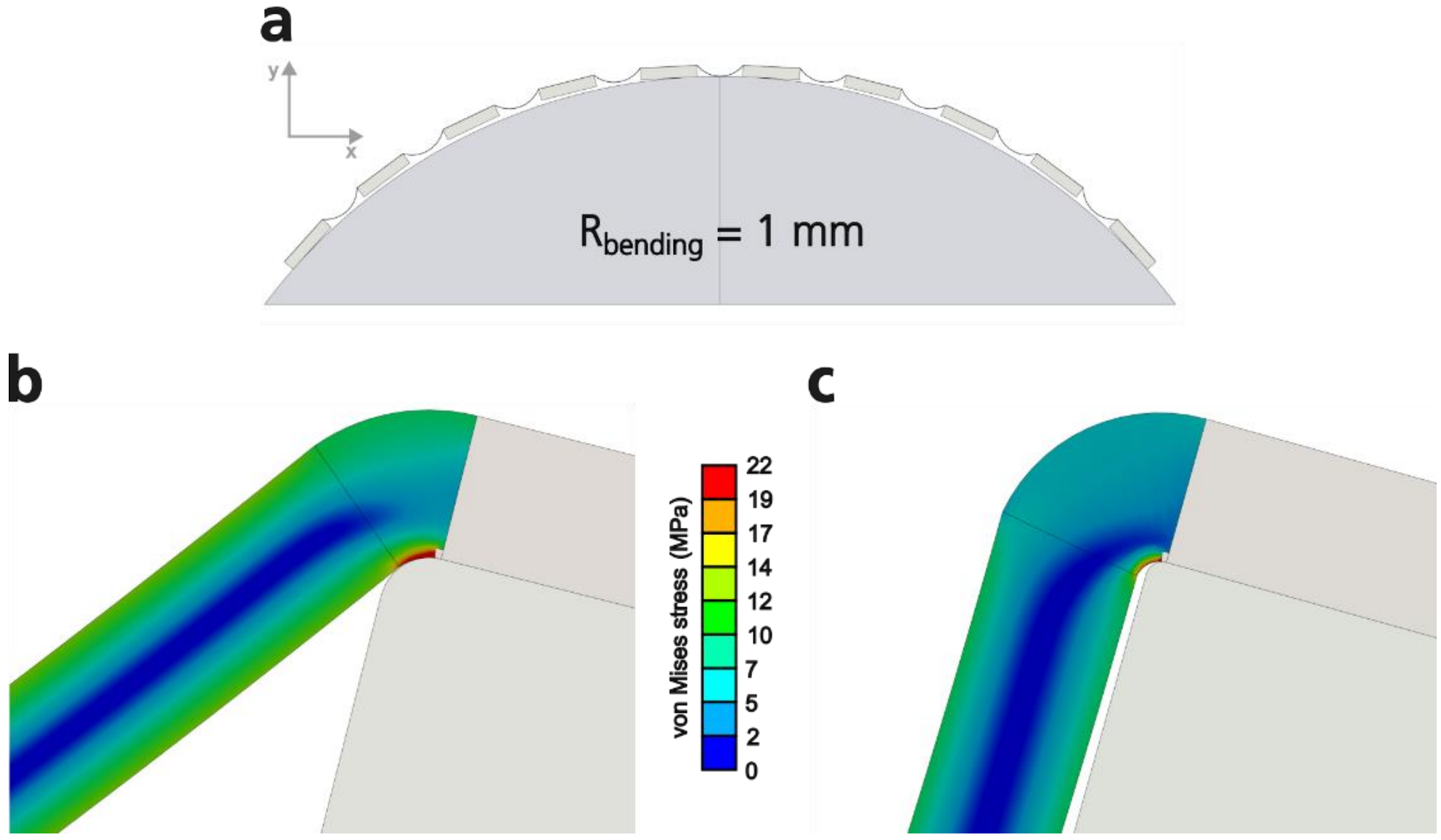

**Figure 3.** a) Schematic of the bending with $R_{bending}$ = 1 mm as the smallest radius. b) Stress distribution of $\sigma_{VM}$ at the transition of the metal to the island during bending with a radius of 1 mm for the geometry with D = 10 µm. c) Stress distribution for geometry with D = 20 µm.

## 2.2 Verification of the Manufacturing Route including Active Devices

To show the feasibility of our three-dimensional island-bridge concept in combination with building up active devices, we used the previous developed core modules, which are shown in **Figure 1a**, and built up a complete manufacturing process where TFT devices on the islands are integrated. Furthermore, these substrates will be used to demonstrate the operation of the TFTs during bending.

**Figure 4a** shows a schematic cross section of the produced island bridge substrate. The polyimide serves both as the device passivation layer on top of the islands and as their encapsulation. As can be seen, the islands in the final substrate are physically separated from one another, so that the polyimide acts as a frame that holds them in a matrix arrangement. The metal bridge is freestanding between each bridge to connect the source, drain and gate contacts.

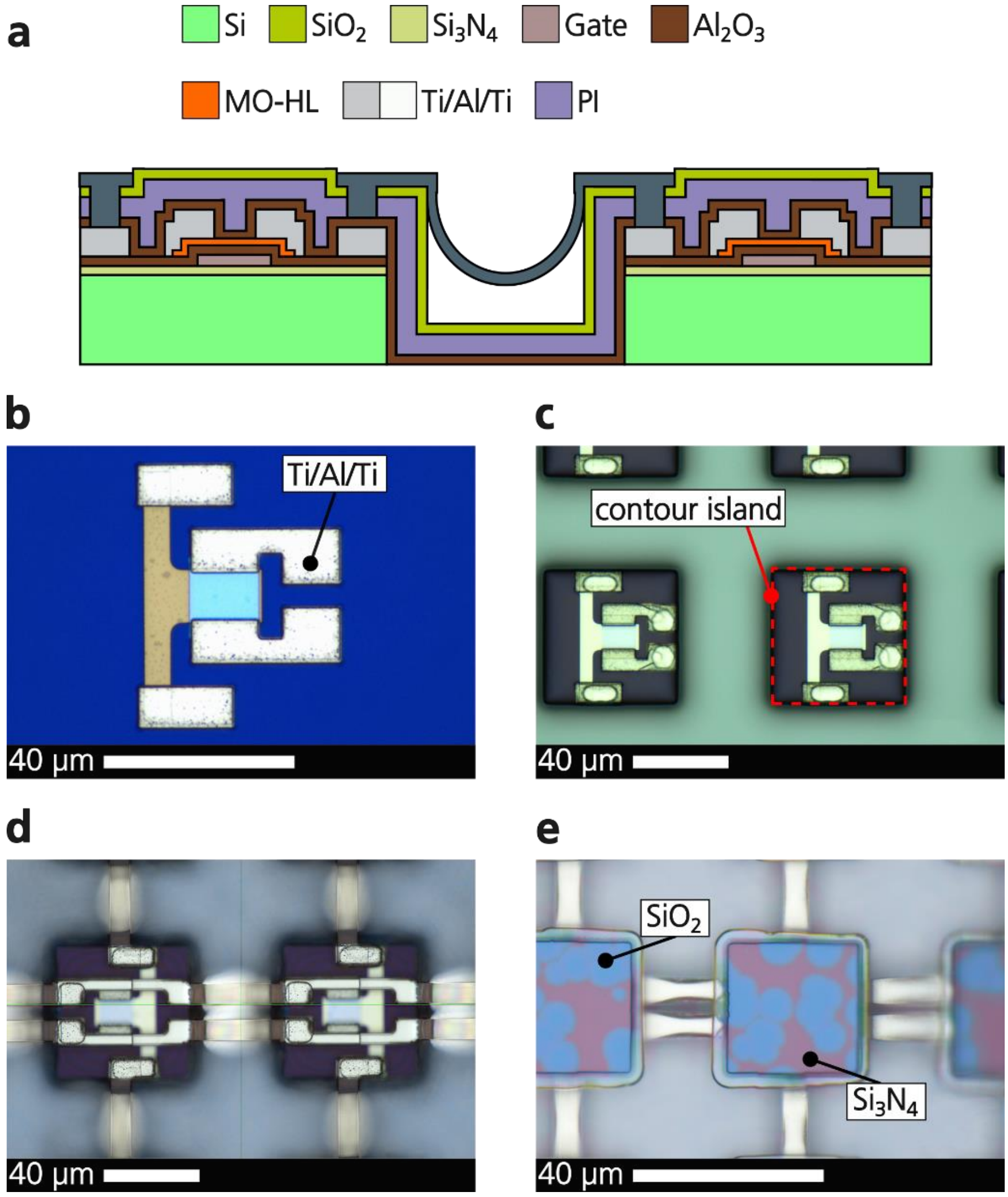

**Figure 4.** a) Schematic cross section of the manufactured island-bridge substrates. b) TFT device fabricated with standard thin-film technology. c) Islands after structuring and polyimide encapsulation. d) Finished front side with 3D bridges acting as metal interconnects. e) Finished backside after bulk silicon removal by KOH.

**Figures 4b–4e** show the processing sequence of the core modules, extending the prior TFT fabrication based on standard thin-film technology. After trench etching to a defined depth, the polyimide encapsulation is formed on the wafer. After the patterning of the polyimide, the polymer is spin coated into the trenches to produce the arc-shaped substructure within the trench depth $D$, which then acts like the fundament for the subsequent metallization. To realize freestanding, three-dimensional bridges between the islands, the arc-shaped trench filling is removed after metal-line fabrication. To achieve the mechanical flexibility of the substrates, the silicon on the backside is removed until the bottom of the trenches is reached separating the islands physically. The island size is 50 µm × 50 µm with a trench width of 40 µm.

### 2.3. Electrical Characterization of manufactured TFT-Devices within the Island-Bridge Structures

#### 2.3.1. TFT Device on Island during mechanical Load

The manufactured substrates, including the active devices described in the previous section, were used to investigate the bending behavior of the structures. In addition, the feasibility of operation of the TFTs under mechanical load was assessed, noting that the brittle silicon islands isolate the devices from tensile strain. Threshold voltage $V_{th}$, subthreshold swing $S$, and saturation mobility $\mu_{sat}$ were used as the critical bending sensitive parameters.[26,27]

**Figure 5a** shows a representative TFT device used for measurements. For this device, the channel length and channel width are $L$ = 10 µm, $W$ = 10 µm ($W$10/$L$10), respectively. To evaluate robustness of the bridges under bending, the measurements of two devices were performed, where metal structures with a width of 30 µm and 10 µm were used. The TFT on the island connected by a bridge width of 10 µm had a channel length and width of 5 µm ($W$5/$L$5). **Figure 5b** presents images of the measurement setup at each bending angle. The 1 mm bending radius was studied only in simulation, as it was not feasible experimentally due to setup limitations. As can be seen, the bending supports for the substrates become progressively higher in order to achieve the corresponding bending radii. Initially, this variation in height could be compensated for by means of the height-adjustable table and the adjustable measuring needles. At $R$ = 2 mm, however, the distance between the measuring needles and the measuring table reached its limit, so that a higher support for $R$ = 1 mm could no longer be accommodated.

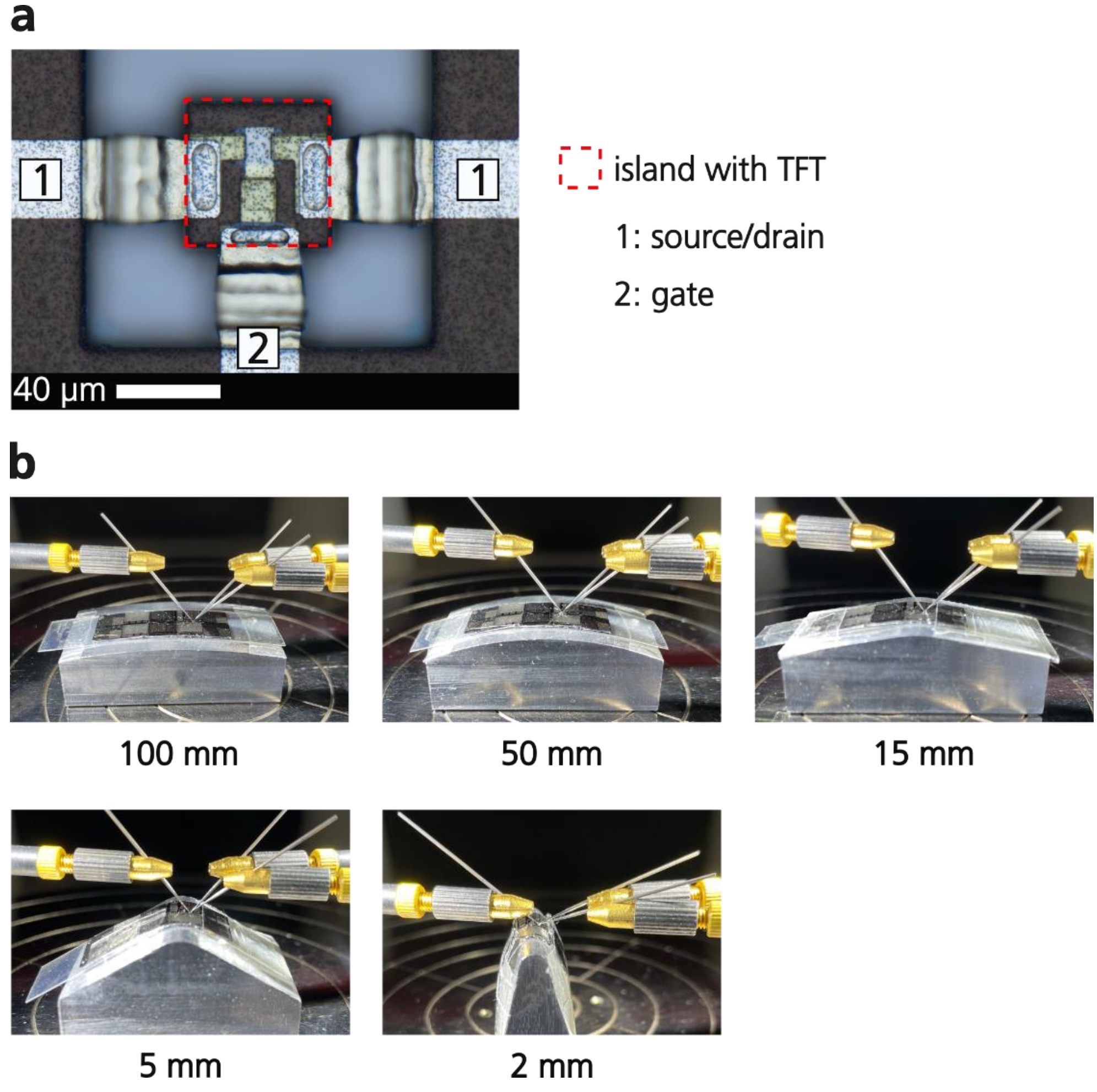

**Figure 5.** a) Example device measured at each bending angle. The bridges to the island have a width of 30 mm and the TFT has a channel length and width of 10 µm. b) Measurement setup for each bending angle.

Using these bending angles, transfer characteristics of the two described TFT devices with different bridge width of 10 and 30 µm were acquired. The results and extracted figures of merit are shown in **Figure 6**. One indicator of the TFT's switching behavior is the on/off current ratio, defined as the ratio of the maximum on-state current to the off-state current ($I_{on}/I_{off}$ ratio). The *W5/L5* device maintains a constant $I_{on}/I_{off}$ at $R$ = 100 mm, 50 mm, and 15 mm. At $R$ = 5 mm the ratio begins to degrade, and the device shows no switching behavior at $R$ = 2 mm, indicating a possible defect. Across radii, $V_{th}$ fluctuates, S remains relatively constant, and at $R$ = 5 mm $\mu_{sat}$ drops to approximately one third of the value measured at a bending radius of 15 mm, reflecting mobility degradation. The switching behavior, meaning the transition from the off state to the on state and thus $V_{th}$, remains within the range of prior measurements. So, one possible reason would be a progressive crack in one of the *S/D* metal bridges, which constricts the conducting path and reduces current flow. Furthermore, an increase in off-current is observed, which may indicate a degradation of the TFT.

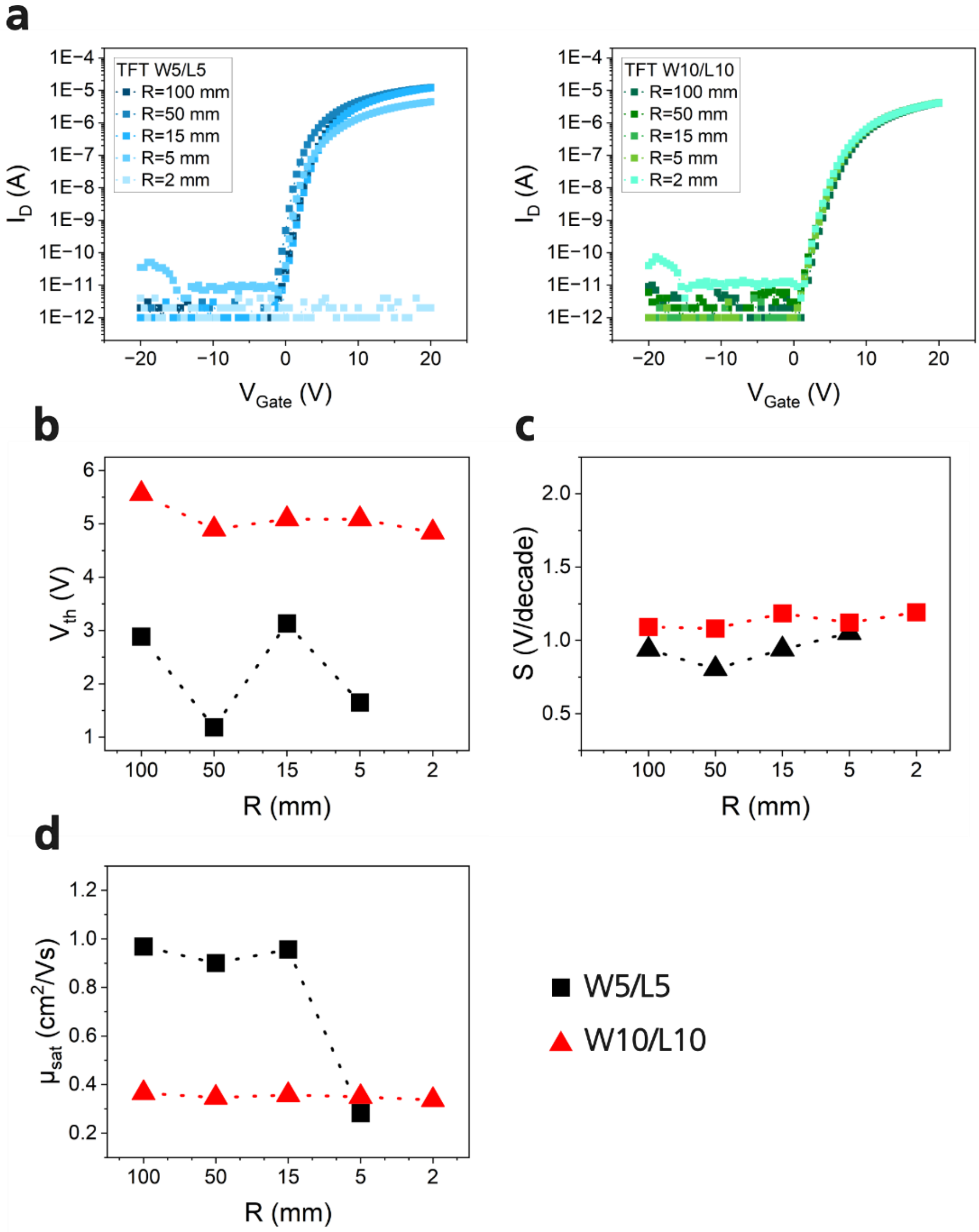

**Figure 6.** Electrical characterization of devices with W = 10 µm, L = 10 µm and W = 5 µm, L = 5 µm, under bending radii from 100 mm to 2 mm. a) Transfer curves for $V_G$ = −20 to 20 V at $V_{SD}$ = 2 V. b–d) Extracted $V_{th}$, S, and $\mu_{sat}$ values.

The *W10/L10* device shows a switching behavior at all bending radii. As expected from the longer channel length the maximum current decreases slightly but remains largely constant across all measurements. The $I_{on}/I_{off}$ ratio decreases only for a 2 mm bend due to higher off-current, resembling the behavior of the *W5/L5* device at $R$ = 5 mm. Across $V_{th}$, $S$, and $\mu_{sat}$ for *W10/L10*, values stay relatively constant, with larger $V_{th}$ and smaller $\mu_{sat}$ compared to the *W5/L5* device. In general, the overall performance of *W10/L10* and of *W5/L5*, at least in the range from 100 to 5 mm, suggests that bending-induced mechanical stress has no pronounced effect on the island and thus device performance. Broader metal bridges appear more robust under bending, while the island acts as a strain-free region for sensitive devices. At least for the metal width of 30 µm this behavior also aligns with numerical simulations, which predicted that even a bending radius of R = 1 mm would cause no critical stress or elongation. Whether a defect was present in one of the bridges of W5/L5 could not be verified after the bending tests, as the structures were too severely damaged while demounting from the supporting surface. As noted, a core concept is that the polyimide serves not only as passivation but also as a flexible substrate. Handling the substrates during testing and mounting them on curved surfaces confirms the polyimide's functionality as a flexible carrier for proper island embedding.

2.3.2. Scaling of the concept

The island-bridge concept was developed as a matrix-like layout, enabling large-area pixelated circuits with high integration density. The precise arrangement integrates seamlessly with the lithographic process of standard planar technology. In a second step, an initial scaling was demonstrated by arranging the individual cells into a 9×9 matrix, as shown in **Figure 7a**. The structures were fabricated following the same route as in **Figure 4**, with a TFT device (*W14/L10*) placed on each island. The arrays are surrounded by contact pads for addressing the gate and source/drain (*S/D*) of each TFT. Typical for multilayer metallization the gate interconnects were isolated through a passivation from the *S/D* metallization.
**Figure 7b** shows a circuit scheme where the gates can be contacted either at the top or at the bottom of the arrays, while source and drain must be connected on the left and right side. The numbering of the columns and rows can be used identifying each device and island. The transfer characteristics of devices in arrays with 10 µm and 5 µm wide bridges (*M10* = Metal width of 10 µm and *M5* = Metal width of 5 µm) where investigated. Measurements were taken on devices 1-1, 2-2 until 9-9 for both variations resulting in the families of curves in **Figure 7c** and **Figure 7d**, respectively. The results show that apart from a single device, all TFTs operate with $I_{on}/I_{off}$ ratios in the 10^5–10^6 range, indicating effective metal-bridge connectivity to each device. Notably, the $I_{off}$ curves for several devices increase linearly from -20 V to $V_{th}$. Within the scope of this study, it was not initially possible to determine the basis for this effect. Further investigations into the structure and the measurement setup are necessary for this purpose. This parasitic behavior complicates subthreshold swing evaluation, since the subthreshold region is not clearly defined. Consequently, only the devices shown in **Figure 7c** were used to evaluate the subthreshold swing *S*. Overall, the results demonstrate successful scaling of the concept and fabrication route into a first matrix arrangement, where active devices at each position are addressable.

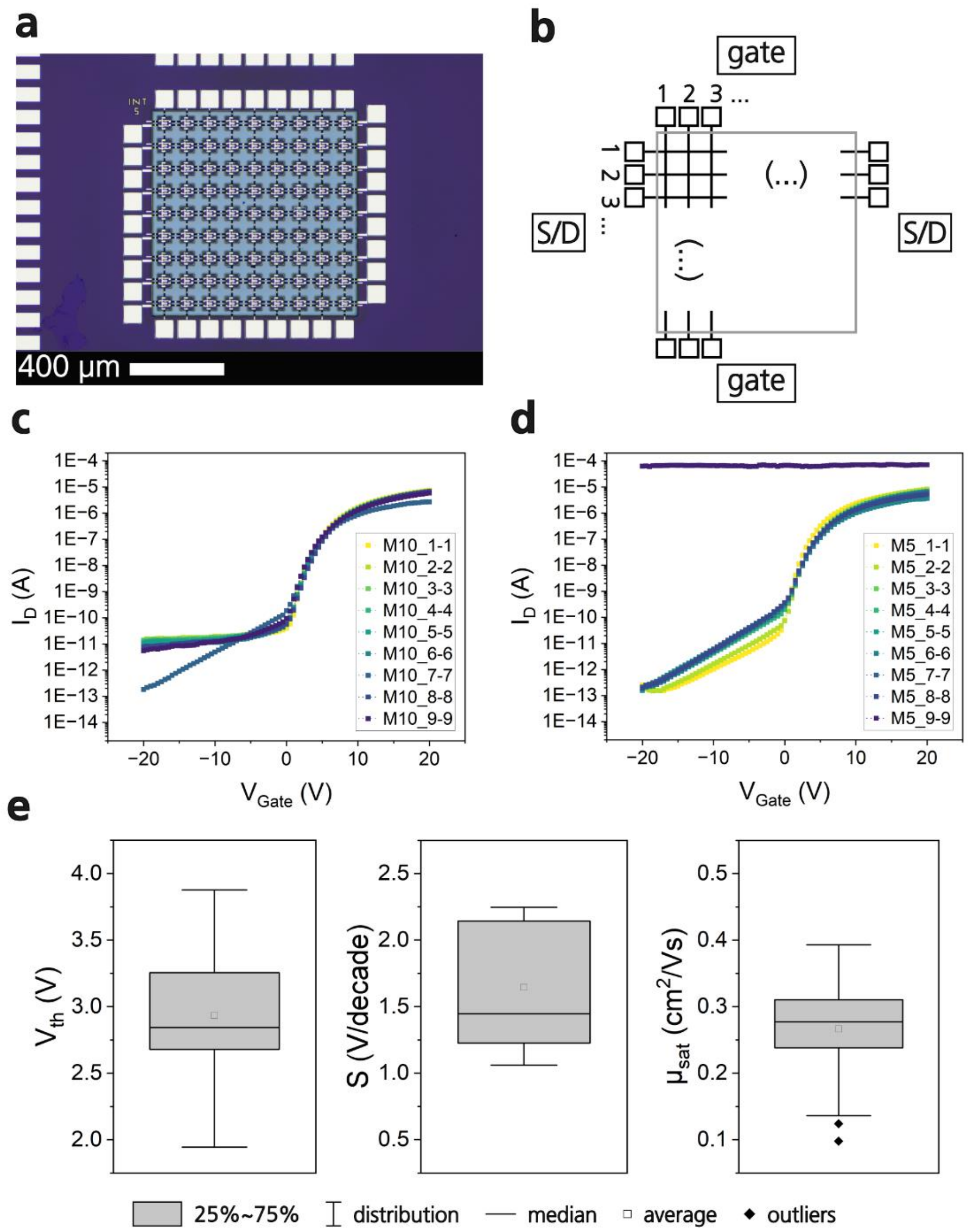

**Figure 7** a) Microscopic image of 9×9 island-bridge matrix with a TFT device on every island after processing. b) Simplified circuit schematic showing gate and S/D connections. c) and d) Transfer characteristics for two matrices, with bridge widths of 10 µm (M10) and 5 µm (M5), respectively. e) Evaluation of $V_{th}$, S, and $\mu_{sat}$ for the matrix with a 10 µm bridge width.

In addition to *S*, **Figure 7e** also presents $V_{th}$ and $\mu_{sat}$. Compared to the single devices measured during bending, $V_{th}$ and $\mu_{sat}$ lie in a similar range. $V_{th}$ is between the *W5/L5* and *W10/L10* values, while $\mu_{sat}$ is closer to the larger-device values. The wider distribution of $V_{th}$ and the spread in *S* (25%–75% percentile) compared to the single devices used for bending test indicates device-to-device variations, which may arise from parasitic effects or process-induced variations. Around 0 V, the TFTs exhibit no abrupt transition, i.e., the current does not increase steeply, which distorts the extraction of S in this region. As can be seen in the supporting information, *S* depends on the slope of the $log_{10}I_D$ curve in the range of -2 to 2 V.

## 3. Conclusion

In summary, we applied our developed CMOS-compatible island-bridge concept with stretch-free, three-dimensional metal bridges and integrated the manufacturing of active devices into the process. This verifies the manufacturability of the route within silicon planar technology and demonstrates that the bridge geometry withstands mechanical stress while the islands act as strain-isolated regions for the devices.

Numerical simulation indicates that, for bending radii down to 1 mm, none of the structures reaches a critical $\sigma_{VM}$ or $\varepsilon_{pl}$. The bridges exhibit the characteristic bending stress distribution, with a neutral plane forming at mid-span and the maximum stress localized mainly at the transitions to the islands. A comparison of trench depth $D$ shows a more favorable stress distribution at larger $D$, identifying $D$ as a key design and process parameter. Plastic strain is not observed until a bending radius of 1 mm, indicating high bending-resistance of the structures.

Experimental investigation demonstrates that TFT devices on the islands remain functional during bending tests, as indicated by relatively stable $V_{th}$, $S$, and $\mu_{sat}$ across the tested radii. A 10 µm bridge (*W5/L5*) degrades starting at $R$ = 5 mm, whereas a bridge with a width of 30 µm exhibits consistent performance across all bending tests, suggesting greater robustness to deformation. Degradation of the *W5/L5* device may arise from device- or bridge-defects. Reproducible bending also confirms the functionality of the polyimide, serving as both passivation and flexible substrate.

The scalability to a 9×9 matrix demonstrates addressability of all TFTs via gate and *S/D* lines and confirms compatibility of the fabrication route with established planar technologies. Parasitic effects are observed in diagonal measurements and must be investigated in further studies.

## 4. Experimental Section

### 4.1. Numerical Simulations

The numerical simulations were performed with ANSYS Mechanical 2024 R1. Simulations were two-dimensional and axisymmetric with respect to the y-axis, thus only half of the geometry was modeled. **Figure 2a** illustrates the configuration with the symmetry boundary. Starting from a half-bridge, five islands were arranged in a row and connected by the corresponding arc-shaped interconnects. To ensure a smooth and realistic transition from metal to island, the edges were rounded. The bending support defined the bending radius. For the support, the ANSYS database provided material properties for a typical steel, used here, while the islands are silicon. A literature-based material model was adopted for silicon.[28] For the aluminum bridges, literature-based parameters were used. Due to the thin geometry of the bridges, precise properties for a thin aluminum film are challenging to obtain. The material model was therefore derived from different publications, with details summarized in the Supporting Information.[29-31] The materials used and their mechanical behavior are summarized in **Table S1**. To simulate bending, the support was fixed and a radial displacement was applied at the last island, with the structure constrained in all directions along the axis of symmetry to enforce radial motion toward the support. By varying the rounded support radius, all results were obtained. Prior to the simulations a mesh-convergence study was performed to ensure accurate results. The details are provided in the Supporting Information (**Figure S1**).

### 4.2. Manufacturing of the Test Structures

Prior to implementing the island-bridge fabrication route, bottom-gate IGZO TFT devices were fabricated on silicon-on-insulator (SOI) substrates. The handling wafer is 300 µm thick, and the device layer is 10 µm, separated by a 1 µm $SiO_2$ layer. An LPCVD deposition of 50 nm $Si_3N_4$ (VF-1000LP, Koyo) provided a backside hard mask at the end of processing. A 50 nm *Ti* layer followed by 10 nm *Al* was deposited by RF magnetron sputtering (CS900S, VonArdenne) and patterned by wet chemical etching to produce the gate electrode. *Al* was etched in an aqueous *HNO3/HF* solution, and *Ti* in *HNO3/H3PO4*. The Gate oxide consisted of 50 nm $Al_2O_3$ deposited by ALD (FlexAL, Oxford Instruments) using trimethylaluminum and $O_2$ as precursors. *IGZO* ($In_2O_3$:$Ga_2O_3$:$ZnO$ = 1:1:1) semiconductor layer was manufactured via RF magnetron sputtering (CS900S, VonArdenne) and patterned in diluted *HCl*. A contact hole for the gate was etched into $Al_2O_3$ in $HNO_3/H_3PO_4$. Source/drain contacts were defined by lift-off after electron beam evaporation of 20 nm *Ti*, 380 nm *Al* and 100 nm *Ti* (Classic 570, Pfeiffer). After TFT fabrication, island structuring into the silicon surface was performed by BOSCH etching ($SF_6$ as etchant, $C_4F_8$ as passivation). A trench depth of 10 µm was reached after 17–20 etch cycles, corresponding to the total device-layer thickness of the SOI substrate. For polyimide passivation, another 50 nm $Al_2O_3$ was deposited by ALD. Polyimide (Pyralin PI 2611, HDMicroSystems) was spin-coated at 3000 rpm (Süss Microtech LabSpin6 TT) and cured on a hotplate (PZ35EB, Harry Gestigkeit) at 300 °C for 1 h. PECVD $SiO_2$ (100 nm) was deposited as passivation (ICP SI 500, Sentech). A contact hole in $SiO_2/PI$ was defined by dry etching ($CF_4$). The underlying $Al_2O_3$ was etched in $HNO_3/H_3PO_4$. For arc-shaped trench filling AZ5214 photoresist (MicroChemicals) was used and spin-coated at 2000 rpm, followed by 500 nm *Al* deposition by sputtering and dry etching (*Cl/HBr*) to define bridges. Finally, resist cleaning employed ashing (TEPLA) with $O_2$ plasma at 300 W for 10 min and an *acetone/DMSO*-based remover at 60 °C, then DI water rinse. Wet backside etching with *KOH* released the freestanding structures; $SF_6$ dry etching was used to pattern the $Si_3N_4$ hard mask, with a single-sided etching frame (AMMT) protecting the front side. The manufacturing was controlled by using optical microscopy (Eclipse L200ND, Nikon) and scanning electron microscopy (JSM-7610F, JEOL).

### 4.3. Electrical Charakterization

The electrical measurements were performed with a probe station from Karl Süss and the Analyzer 4156A from Hewlett Packard. Gate voltage $V_G$ was swept from −20 V to +20 V in 0.5 V steps, while the drain current $I_D$ was recorded. The source–drain bias was fixed at 2 V, yielding the TFT transfer characteristics used to extract the device parameters. Using these parameters, measurements were performed for both isolated bending tests at various bending radii and for 9×9 island-bridge matrices. The measurement methodology was the same for all measurements.

For bending radii of 100, 50, 15, 5, and 2 mm, rectangular aluminum supports with a curved cross-section and thus to provide the respective radii were fabricated to enable stable and accurate mounting of the substrates during measurement.

$V_{th}$ and $S$ were obtained from the numerical data of $\sqrt{I_D}$ versus $\sqrt{I_G}$ and $log_{10}I_D$ versus $I_G$, respectively (procedure shown in Supporting Information). The field-effect mobility $\mu_{sat}$ was computed from **Equation S1** in the Supporting Information.

**Acknowledgements**

Fundamental developments related to the work presented here were supported by the Fraunhofer Internal Programs under Grant No. MAVO 840 092. The authors also want to acknowledge the Chair of Electron Devices (LEB) at the Friedrich-Alexander-University Erlangen-Nuremberg (FAU) and its Prof. Dr.-Ing. Jörg Schulze, as well as the technical staff for the support and providing access to the laboratories. Thanks also to Dr. Norman Böttcher and Dr. Mathias Rommel for their support.

**Conflict of Interest**

The authors declare no conflict of interest.

**Data Availability Statement**

The data that support the findings of this study are available from the corresponding author upon reasonable request.

# Supporting Information

**Stretch-free, shape-induced 3D island–bridge networks for flexible TFTs on silicon planar technology verified through bending and scalability to 9×9 matrix**

*Daniel Joch*, Robert Kammel, Fabian Magerl and Michael P.M. Jank*

## 1. Numerical Simulations

A material model that diverges from bulk aluminum is required due to the distinct material properties of thin films, predominantly governed by the decreasing grain count across the cross-section. Fewer grains also imply a reduced number of potential dislocation paths, thereby limiting plastic deformation.[29, 32-34] Soare et al. investigated the material properties of aluminum using nanoindentation tests on thin films, so that we integrated these findings into a material model within the ANSYS software.[29] Additionally, following Zhao et al., aluminum is assumed to exhibit ideal plastic behavior.[30] Further literature shows the investigation of the hardening behavior of aluminum up to 1% plastic strain $\varepsilon_{pl}$. It can therefore be assumed that in this range no fracture occurs, but rather only hardening takes place in the material, which in turn can increase $\gamma_{ys}$.[31]
All materials employed in numerical modelling and their properties are catalogued in **Table S1**.

**Table S1.** Summary of the materials and their mechanical properties for the use in the numerical modelling.

| Bridge | Insel | Support |
|---|---|---|
| Aluminum<br>isotropic elastic<br>$E^{a)}$ = 73 GPa, $v^{b)}$ = 0.33<br>bilinear isotropic hardening<br>$\gamma_s^{c)}$ = 545 MPa<br>$E_t^{d)}$= 2.5 GPa | Silicon (100)[28]<br>Isotropic elastic<br>$E$ = 130 GPa, $v$ = 0.29 | Steel (Ansys database)<br>Isotropic elastic<br>$E$ = 200 GPa, $v$ = 0.3 |

[a)] Young's modulus; [b)] Poisson's ratio; [c)] yield strength; [d)] tangent modulus

To prepare for the numerical simulations, a convergence study was conducted to generate a sufficiently fine mesh, with the aim of ensuring that the results for the parameters considered are as accurate as possible. In our assessment, we examined the von Mises stress $\sigma_{VM}$ in dependence of the mesh element size as it is the parameter we used for our investigations. Additionally, potential energy ($E_P$) minimization was evaluated as another indicator.[35,36] Depending on the energy, a FEM simulation calculates the deformation and, through minimization, determines the state of equilibrium for the system.[37] Furthermore, a comparison was made between the averaged and non-averaged von Mises stress values. For the averaged values, neighbor element contributions were included in the calculation; for the non-averaged values, only the nodal value of each element was used.[38] The element size was explored over a range from 0.3 to 0.01 µm and the results for $\sigma_{VM}$ and $E_P$ are normalized to the values corresponding to the smallest element size since these are considered the most accurate ones. Quadratic and triangular mesh elements were compared due to simulation efficiency. The results are presented in **Figure S1** as a plot of standard stress and standard energy versus the number of elements.

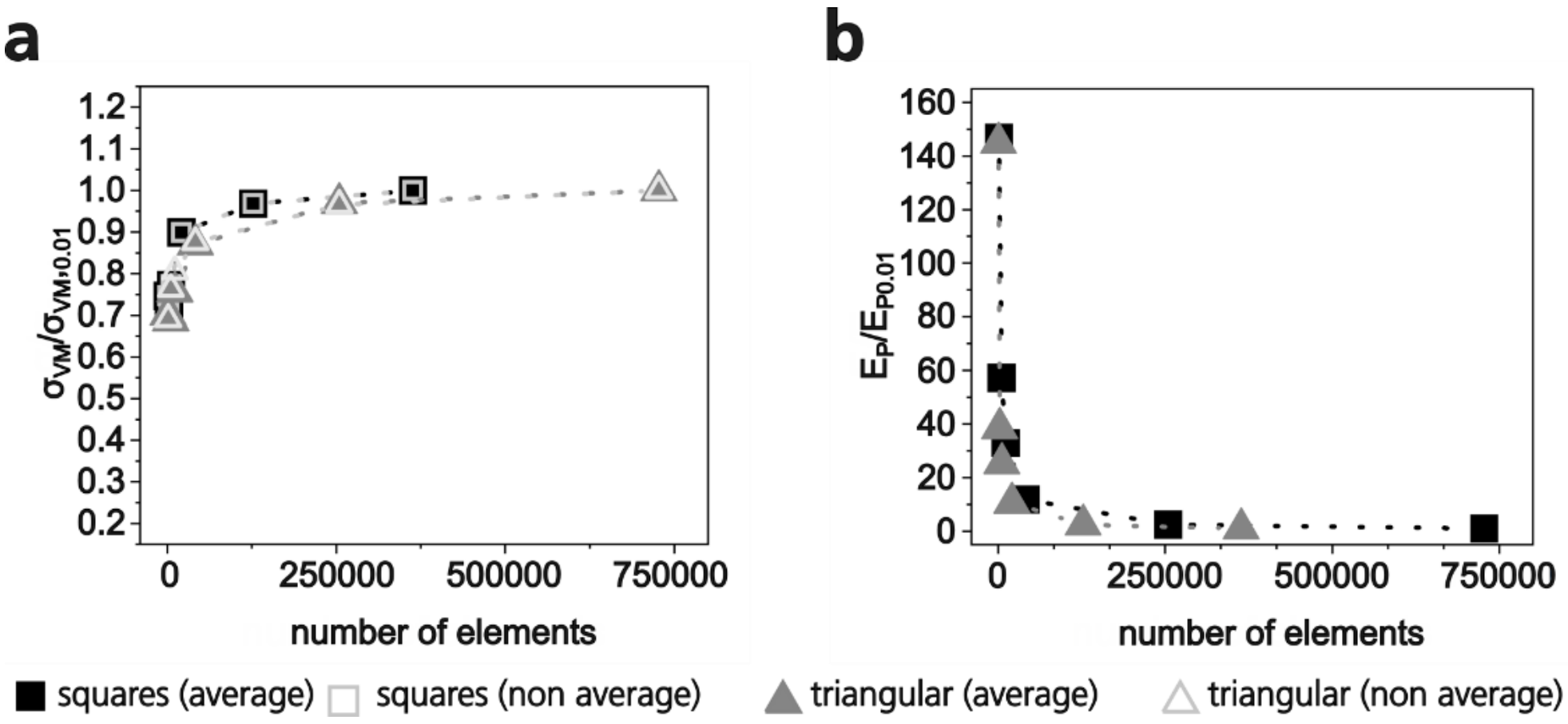

**Figure S1**. a) Evaluation of mesh size for von Mises stress in the bridge. The stress σ is normalized to the value $σ_{0.01}$. B) evaluation of the mesh size for potential energy E, also normalized to $E_{0.01}$.

As can be seen in **Figure S1 a)** sufficiently small convergence deviation can be reached with increasing the number of square and triangle elements. A similar trend can also be observed in **Figure S1 b)** for the minimization of potential energy. Moreover, starting at an element size of 0.03 µm, there is no longer any deviation between averaged and non-averaged values. From that a total of 127,275 elements were identified, corresponding to an element size of 0.02 µm, and thus was adopted for the size during meshing. Square elements were chosen because they achieve a similar level of convergence with fewer elements and are therefore more efficient. **Figure S2** shows the final mesh quality.

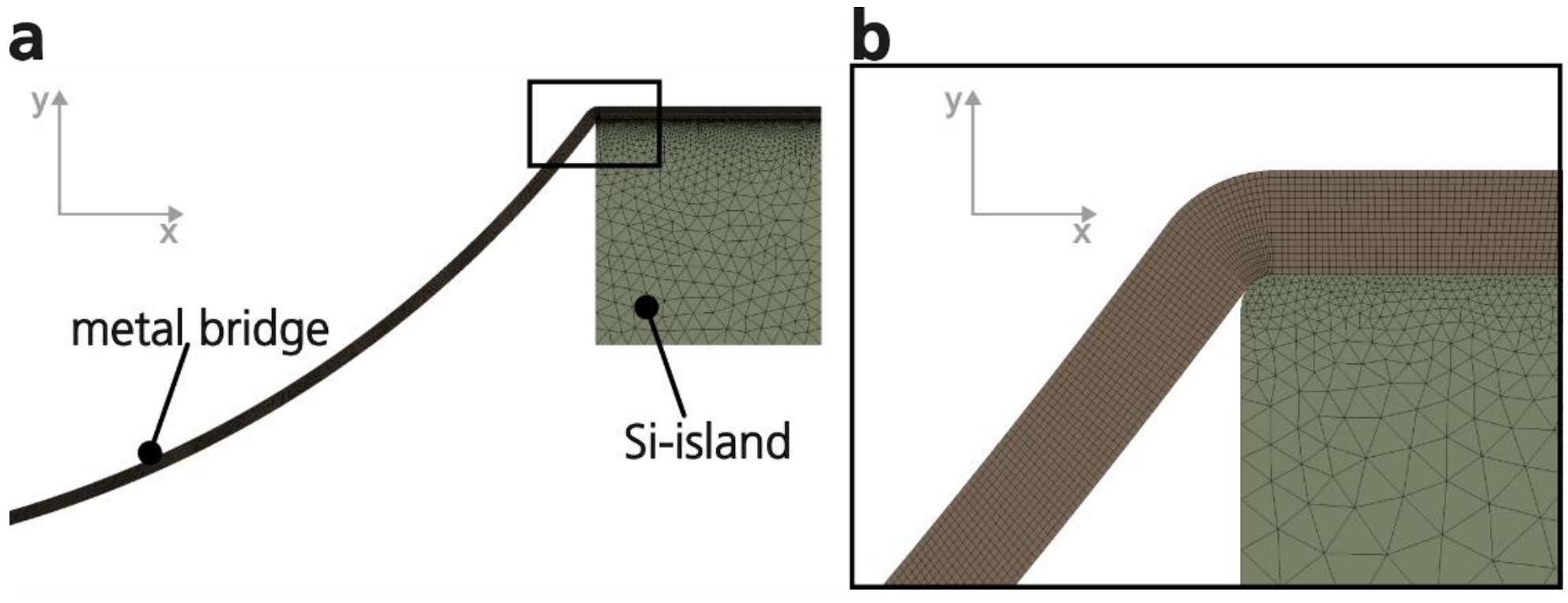

**Figure S2**. Final Meshing with an element size of 0.02 µm for the part of the metal bridge. a) half of the bridge with beginning of silicon island. b) detailed view of transition on the silicon island.

## 2. Electrical Characterization

For the evaluation of electrical performance of the TFT devices the parameters threshold voltage $V_{th}$, subthreshold swing $S$ and saturation mobility $\mu_{sat}$ were extracted and calculated, respectively. $V_{th}$ and $S$ were obtained from data of $\sqrt{I_D}$ versus $\sqrt{I_G}$ and $log_{10}I_D$ versus $I_G$.[39,40] An example of the extraction of both is shown in **Figure S3**.

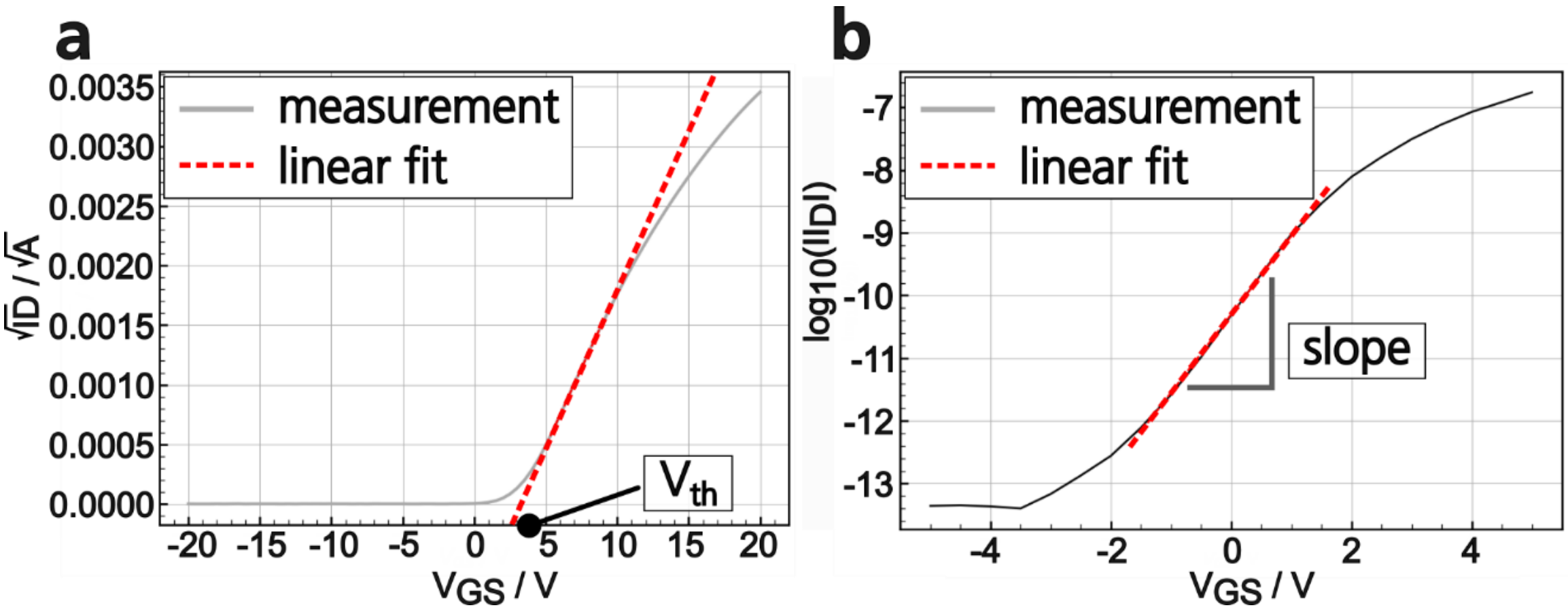

**Figure S3.** Graphical extraction of the TFT characteristics. a) threshold voltage $V_{th}$. b) subthreshold swing *S*.

The parameter $\mu_{sat}$ can be calculated (**S1**) with given values from the device geometry and $C_{OX}$.[39]

$$\mu_{sat} = \frac{2Lm^2}{WC_{OX}} \quad \textbf{(S1)}$$

Here, *L* is the length, and *W* is the width of the channel. The value m is the slope of the fit of $\sqrt{(I_D)}$ versus $V_{GS}$, which is already determined for the extraction of $V_{th}$. $C_{OX}$ is the capacitance of the gate oxide, which is $Al_2O_3$, deposited by atomic layer deposition (FlexAL, Oxford Instruments). For this material deposition, we did an in-house characterization, where we also evaluated $\varepsilon_r$. From this, we were able to determine the value of 8, which consequently was used together with deposited thickness $Al_2O_3$ for the area-based calculation of $C_{OX}$.